\documentclass[aps,twocolumn,groupedaddress,showpacs,nofootinbib]{revtex4}

\usepackage{graphicx}
\usepackage{hyperref}
\usepackage{dcolumn}
\usepackage{bm}
\usepackage{epsfig,amsbsy,bm,amssymb,amsmath}
\usepackage{xcolor}

\renewcommand{\k}{{k}}
\newcommand{\hs}{\hspace*}
\newcommand{\vs}{\vspace*}

\newcommand{\w}{\omega}
\newcommand{\W}{\Omega}

\newcommand{\eref}[1] {(\ref{#1})}
\newcommand{\Eref}[1] {Eq.~(\ref{#1})}
\newcommand{\Fref}[1] {Fig. \ref{#1}}

\newcommand{\ra}{\rangle}
\newcommand{\la}{\langle}
\newcommand{\nn}{\nonumber}
\newcommand{\be}{\begin{equation}}
\newcommand{\ee}{\end{equation}}
\newcommand{\br}{\begin{eqnarray*}}
\newcommand{\er}{\end{eqnarray*}}
\newcommand{\ba}{\begin{eqnarray}}
\newcommand{\ea}{\end{eqnarray}}
\newcommand{\bp}{\begin{minipage}}
\newcommand{\ep}{\end{minipage}}
\newcommand{\bt}{\begin{tabular}}
\newcommand{\et}{\end{tabular}}

\newcommand{\ms}{\vspace*{-5mm}}
\newcommand{\mms}{\vspace*{-2.5mm}}
\renewcommand{\l}{\lambda}

\renewcommand{\k}{{\bm k}}

  \newcommand{\q}{{\bm q}}

\renewcommand{\l}{\lambda}

\renewcommand{\q}{\theta}

\renewcommand{\q}{\theta}

\renewcommand{\l}{\lambda}

\newcommand{\isum}%
{\mathop{\hbox{$\displaystyle\sum\kern-13.2pt\int\kern1.5pt$}}}

\newcommand{\Wcm}[2]{
$\rm {#1}\times10^{{#2}}~W/cm^2$}

\begin{document}
\bibliographystyle{apsrev}
%



\title{Extraction of XUV+IR ionization amplitudes 
from the  circular dichroic phase}

\author{Anatoli~S. Kheifets}

\affiliation{Research School of Physics, The Australian National
  University, Canberra ACT 2601, Australia}

 \date{\today}


\begin{abstract}
A strong helicity dependence of reconstruction of attosecond bursts by
beating of two-photon transitions (RABBITT) with circularly polarized
XUV and IR pulses was reported by Han {\em~et~al.} [Nature Physics 19,
  230 and arXiv 2302.04137 (2023)]. They attributed a circular
dichroic phase in RABBITT to the helical structure of the
photoelectron wave packets in the final state. We exploit this effect
to determine the magnitude and phase of two-photon XUV+IR ionization
amplitudes. In $s$-electron targets (H, He, Li), such a determination
is fully {\em ab initio} and free from any instrumental
limitations. In heavier noble gases like Ar, characterization of
two-photon ionization amplitudes can be made from the circular
dichroic phase with minimal and very realistic assumptions.
\end{abstract}

\maketitle

Circular dichroism (CD) in atomic and molecular photoionization has
been studied intensely in recent years. In chiral molecules, it is
attributed to the left and right handedness of molecular enantiomers
\cite{Cireasa2015}. In atoms, electron ring current in various
magnetic sublevels can be co- or counter-rotating (CO or CR) with the
circularly polarized ionizing radiation. This results in different
ionization probabilities \cite{Barth2011,Barth2013,Walker2021} and
time delays \cite{PhysRevA.87.033407}. The CD effect is inherent in
atomic double photoionization where the momenta of the two receding
photoelectrons and the propagation direction of the photon form a
chiral triangle
\cite{Berakdar1992,Viefhaus1996,Mergel1998,KB98e}. Chirality can be
also imprinted on atoms by a synthetic chiral light \cite{Mayer2022}.
Very recently, the CD effect has been observed in the process of
reconstruction of attosecond bursts by beating of two-photon
transitions (RABBITT) driven by circularly polarized radiation
\cite{HanMeng2023,Han2023Separation}. Depending on the co- and
counter-rotating of the extreme ultraviolet (XUV) pump and the
infrared (IR) probe pulses, the atoms exhibited different set of
RABBITT magnitude and phase parameters.  This effect was attributed to
the helical structure of the photoelectron wave packets in the final
state.  While in the CR case, absorption of an IR photon leads to the
final state composed of two partial waves, the CO configuration
corresponds to the final state with just a single partial wave. This
disparity is responsible for the RABBITT CD effect.  The same single
and dual wave disparity creates a circular dichroic time delay in
single XUV photon ionization \cite{PhysRevA.87.033407}. However, its
observation requires a polarized target atom while in RABBITT such a
polarization is not necessary. The RABBITT process with linear
polarization is driven by interference of the two dual wave final
states and their amplitudes and phases are entangled.  Disparity of
the circular polarization RABBITT creates an opportunity to
disentangle the two interfering final states and to extract the
corresponding ionization amplitudes and phases individually. In the
present work, we demonstrate such an extraction for the helium atom
and show that it is free from any instrumental limitations and
inaccuracies. This way a complete experiment can be performed in
two-photon XUV+IR ionization similarly to single XUV photon ionization
\cite{Shigemasa1998,Wang2000}. In heavier noble gases like argon, the
full characterization of the two-photon ionization amplitudes can be
made from the circular dichroic phase using minimal and very realistic
assumptions.

In a RABBITT measurement \cite{PaulScience2001,Mairesse1540}, an
ionizing XUV attosecond pulse train (APT) is superimposed on an
attenuated and variably delayed replica of the driving IR pulse.  The
XUV photon $\W^\pm=(2q\pm1)\w$ is absorbed from the initial bound
state and then is augmented by an IR photon absorption $+\w$ or
emission $-\w$ leading to formation of the even order sideband (SB) in
the photoelectron spectrum.  The center of the IR pulse is shifted
relative to the APT by a variable delay $\tau$ such that the magnitude
of a SB peak oscillates as
\be
S_{2q}(\tau) =
A+B\cos[2\omega\tau-C]
\ .
\label{oscillation}
\ee
The RABBITT parameters $A$, $B$ and $C$ entering
Eq.~(\ref{oscillation}) can be expressed as 
\ba
\nn
A&=&\sum_{m}|{\cal M}^{(-)}_{m}(\k)|^2+|{\cal M}^{(+)*}_{m}(\k)|^2
\\
\nn
B&=&2{\rm Re}\sum_{m} \left[{\cal M}^{(-)}_{m}(\k)
{\cal M}^{(+)*}_{m}(\k)\right]
\\
C&=& \arg\sum_{m}\left[
{\cal M}^{(\rm -)}_{m}(\k)
{\cal M}^{(\rm +)*}_{m}(\k)
\right]
\equiv 2\w\tau_a
\ .
\label{abc}
\ea
Here ${\cal M}^{(\pm)}_{m}(\k)$ are complex and angle-dependent
amplitudes of two-photon ionization produced by adding $(+)$ or
subtracting $(-)$ an IR photon, respectively. An incoherent summation
over the angular momentum projection of the initial state $m$ is
explicit in \Eref{abc}.  The atomic time delay $\tau_a$ quantifies the
timing of the XUV ionization process.

The angular dependence of the amplitudes ${\cal M}^{\pm}(\k)$ can be
deduced from an analytic  expression \cite{Dahlstrom201353}:
\ba
\label{lopt}
\nn
{\cal M}^{\pm}_{m}(\k) &\propto&
\sum_{\l=l\pm1}
\sum_{L=\l\pm1}
\sum_{|M|\leq L ; \mu|\leq\l}
(-i)^L e^{i\eta_L} Y_{LM}(\hat{k})
\\\nn&&\times
\isum \ d^3\kappa \
{\la R_{kL}|r|R_{\kappa\l}\ra \la R_{\kappa\l}|r|R_{ln}\ra\over
  E_i+\W^\pm-\kappa^2/2-i\gamma}
\\
&&\ \ \times  
\la Y_{LM}|Y_{1m_2}|Y_{\l\mu}\ra 
\la Y_{\l\mu}|Y_{1m_1}|Y_{lm}\ra 
\ea
Here $\la nl|, \la \kappa\l|$ and $\la kL|$ are the initial,
intermediate and final electron states defined by their linear and
angular momenta. The linear polarization (the LIN case) corresponds to
$m_1=m_2=0$ whereas for the circular polarization $m_1=1$ and
$m_2=\pm1$ in the CO/CR cases, respectively.

For an $s$-electron target, $l=m=0$ and $\l=1$. The LIN and CR cases
correspond to $M=0$ and $L=\{0,2\}$ whereas the CR case has $M=2$
which excludes $L=0$. 
By carrying out the angular
integration in \Eref{lopt} we can write
\mms
\be
\label{Mk}
{\cal M}^{\pm}(\k) = 
\left\{
\begin{array}{rrc}
T^{\pm}_0 + 2P_2(\cos\q)T^{\pm}_2 & {\rm LIN} \,\, \\
-T^{\pm}_0 +  P_2(\cos\q)T^{\pm}_2 & {\rm CR} \,\,\,\, \\
e^{2i\phi}\sin^2\q \ T^{\pm}_2 & {\rm CO} \, ,\\
\end{array}\right.
\ee
where $T^\pm_L$ absorbs the radial parts of \Eref{lopt}.  In the LIN
case, $\q=90^\circ$ defines the photon propagation direction whereas
in the CO/CR cases this direction corresponds to
$\q=0/180^\circ$. That is why to compare the linear and circular
cases, we shift the CO/CR angular scale by $90^\circ$.  The polar
angle $\phi$ in \Eref{Mk} is immaterial because of the rotational
symmetry of the ionization process. For the same reason, the
directions $\q$ and $180^\circ-\q$ are equivalent. So only half of the
azimuthal angular range needs to be analyzed.

\Eref{Mk}  defines the RABBITT phase for $\phi=0$:
\ba
\label{CC}
C^{\rm LIN} &=& 
 \arg\Big[T_2^{-}T_2^{+*}\Big] +
\\&&\hs{-2cm} \arg\left[2P_2(\cos\q)+{T^{-}_0 \over T^{-}_2}\right]
+ \arg\left[2P_2(\cos\q)+\left({T^{+}_0 \over T^{+}_2}\right)^*\right]
\nn
\\
C^{\rm CR/CO} &=& 
 \arg\Big[T_2^{-}T_2^{+*}\Big]
+ \arg\left[P_2(\cos\q)-{T^{+/-}_0 \over T^{+/-}_2}\right]
\nn
\ea
We observe in \Eref{CC} that the phases of the $\pm$ transition
amplitudes are entangled in the LIN case whereas they stand alone in
the CR/CO cases.  For further analysis, we rewrite the ratio
$
T^\pm_0 / T^\pm_2 = R^\pm \exp(\pm i\Delta\Phi)
$
and note that $R^{+}<1<R^{-}$ by virtue of the \citet{PhysRevA.32.617}
propensity rule for the continuous-continuous (CC) transitions
\cite{Busto2019}. We also use the emission/absorption phase identity
$\Delta\Phi \equiv \Phi_{L'}^{-}-\Phi_{L}^{+} \approx
\Phi_{L'}^{\pm}-\Phi_{L}^{\pm}
$
postulated in \cite{Boll2023}. Based on this identity and \Eref{CC}, 
$
\Delta\Phi = C_{\rm CO}(\q_{\rm m}) - C_{\rm CR}(\q_{\rm m})
$
where the ``magic angle'' defines the node of the Legendre
polynomial $P_2(\q_m)=0$.  With all these observations, we can conclude
that the LIN phase is sandwiched between the CO and CR ones: \mms
\be
\label{less}
C_{\rm CR}(\q)<C_{\rm LIN}(\q)<C_{\rm CO}(\q)
\ \ \forall \q
\mms
\ee
To prove that this relation is indeed satisfied, we conduct a set of
simulations by solving numerically the time-dependent Schr\"odinger
equation which describes the helium atom driven by the RABBITT
configuration of pulses. Numerical details of these simulations can be
found in our recent works
\cite{Kheifets2023,Kheifets2023polarization}. Most essentially, the IR
carrier wavelength is set in the 800~nm range and the XUV and IR field
intensities are kept within \Wcm{1}{10} range. The latter condition
keeps the ionization process within the boundaries of the lowest order
perturbation theory (LOPT) which is used to derive \Eref{lopt}.

\begin{figure}[t]
\epsfxsize=6.cm
\epsffile{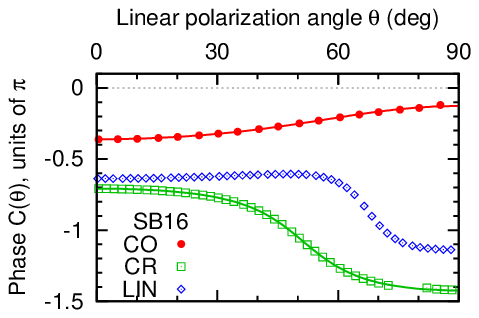}
\vs{-1cm}

\epsfxsize=6.cm
\epsffile{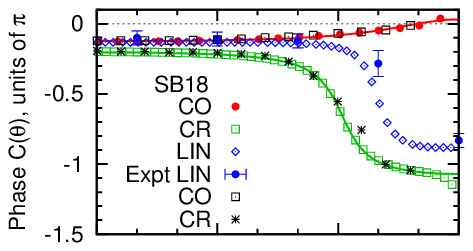}
\vs{-1cm}

\epsfxsize=6.cm
\epsffile{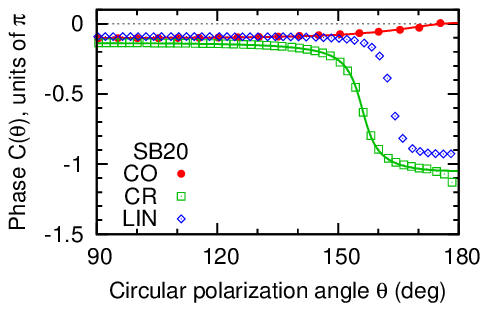}

\caption{Angular dependent RABBITT phase ($C(\q)$ parameter) in
  SB16-20 of helium obtained with CO (red circles) and CR
  (green squares) circular polarization as well as with linear (LIN)
  polarization (blue diamonds). The CO and CR fit with \Eref{CC} is shown
  with red and green solid lines, respectively. The experimental
  linear phase for SB18 is from \cite{Jiang2022} and CO/CR phases are
  from \cite{Han2023Separation}.
\label{Fig1}
\ms
}
\end{figure}

The results of the TDSE calculations on He are exhibited in
\Fref{Fig1} for the lowest sidenabds SB16-20.  Our numerical results
fully support \Eref{less} and are in good agreement with the
experimental LIN phase for SB18 reported in \cite{Jiang2022}.  Another
important observation is that the boundaries of the CO/CR phases that
encompass the LIN phase become narrower as the SB order
grows. Simultaneously, the phase drop by $\sim\pi$ becomes
steeper. The latter observation for the LIN phase has already been
made and attributed to the convergence $R^\pm\to1$
\cite{PhysRevA.94.063409,PhysRevA.96.013408,Boll2023}.

\begin{figure}[h]
\epsfxsize=6.5cm
\epsffile{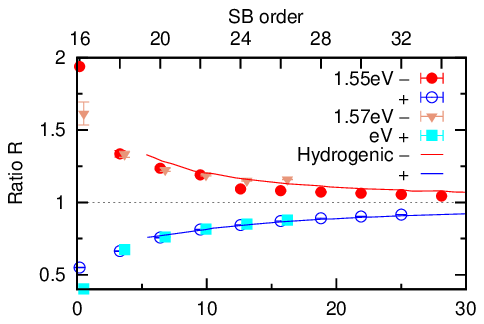}
\vs{-5mm}

\epsfxsize=6.5cm
\epsffile{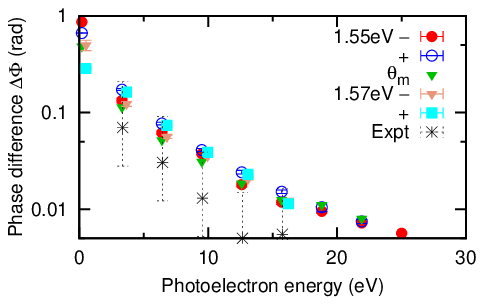}

\caption{Top: ratios $R^\pm=|T_0^\pm/T_2^\pm|$ as extracted from the
  fit of \Eref{CC} to RABBITT phases of helium $C(\q)$ calculated in
  TDSE and shown in \Fref{Fig1}. Comparison is made with a hydrogenic
  model employed in \cite{Boll2023}. Bottom: the phase differences
  $\Delta\Phi^\pm=\arg [T_0^\pm/T_2^\pm]$ as extracted from the same
  fit as well as the magic angle difference $ \Delta\Phi(\q_{\rm
    m})$. The experimental values are from \cite{Han2023Separation}.
\label{Fig2}
\mms}
\end{figure}

While the two  ratios $R^\pm$ enter the LIN phase in \Eref{CC}, the
CO/CR phases contain them separately. Hence, we can fit our
numerical CO/CR phases with the analytic expression \eref{CC} and obtain
the corresponding ratios $R^\pm$ and phase differences $\Delta\Phi^\pm$. 
Results of this procedure are exhibited in \Fref{Fig2}. The top panel
shows the ratios $R^\pm=|T_0^\pm/T_2^\pm|$ whereas the bottom panel
displays the phase differences $\Delta\Phi^\pm=\arg
[T_0^\pm/T_2^\pm]$. Thus extracted ratios $R^\pm$ are compared with
the values returned by a hydrogenic model deployed in \cite{Boll2023}.
As expected $R^\pm\to1$ and $\Delta\Phi^\pm\to0$ as the photoelectron
energy grows. The emisson/absorption phase identity as well as the
phase determination at the magic angle are well supported by our
numerical results.  With our ratio determination, we can estimate an
excess of the CO RABBITT magnitude over the CR one by the value of
+10\% for SB32 and +7\% for SB34 which is very similar to
\cite{HanMeng2023} and contrary to \cite{Mazza2014}. The latter work
claims the CR excess over the CO one.
%

\mms
\ms\ms

The hydrogenic model exploited in \cite{Dahlstrom201353,Boll2023}
looses its validity close to threshold at very low photoelectron
energies. In the present case, this model clearly fails for SB16. This
sideband is formed by an IR photon absorption via an intermediate
discrete bound state.  Such an under-threshold (or uRABBITT) process
has been studied extensively in He
\cite{SwobodaPRL2010,Drescher2022,Neoricic2022,Autuori2022} and in
heavier noble gases - Ne
\cite{Villeneuve2017,PhysRevA.103.L011101,Kheifets2021Atoms} and Ar
\cite{Kheifets2023}. In helium, the discrete phase $\Phi^+$ oscillates
with the IR photon energy $\w$ when the submerged harmonic peak H15
passes through the discrete $1s3p$ level. As an illustration of this
oscillation, we compare in \Fref{Fig2} the two sets of TDSE
calculations performed at the central IR frequency of 1.55~eV and
1.57~eV. Such a minuscule photon energy variation causes a significant
change of the ratio and the phase difference for SB16 while these
parameters for other sidebnads remain barely changed.
\mms

In  $p$-electron targets, the parameterization of the
two-photon amplitudes depends on the orbital momentum projection
$m$. In the CO case, these $m$-specific amplitudes take the form:
\ba
\label{Mp}
{\cal M}_{m=0}^{-} &\propto&
\cos\q \
[-T^{-}_1+  \bar P_3(\cos\q) \ T^{-}_3]  
\\\nn
{\cal M}_{m=0}^{+} &\propto&
e^{2i\phi}\sin^2\q\cos\q \ T_3^{+}
\\\nn
{\cal M}_{m=+1}^{-} &\propto&
\sin\q \
[-6 T^{-}_1  + \bar P_3^1(\cos\q) \ T^{-}_3 ]
\\\nn
{\cal M}_{m=+1}^{+} &\propto&
e^{3i\phi}\sin^3\q \
T^{+}_3
\\\nn
{\cal M}_{m=-1}^{\pm}  &\propto&
\mp
\sin\q \
[- T^{\pm}_1 + \bar P_3^1(\cos\q) \ T^{\pm}_3]
\ea
The CR amplitudes can be obtained by permutation of the
emission/absorption $+/-$ superscripts.  In \Eref{Mp} we introduce $\bar
P_3(\cos\q) = P_3/P_1=(5\cos^2\q-3)/2$ and $\bar P^1_3(\cos\q) =
P^1_3/P^1_1=3(5\cos^2\q-1)/2$. 

\begin{figure}[t]
\epsfxsize=6.5cm
\epsffile{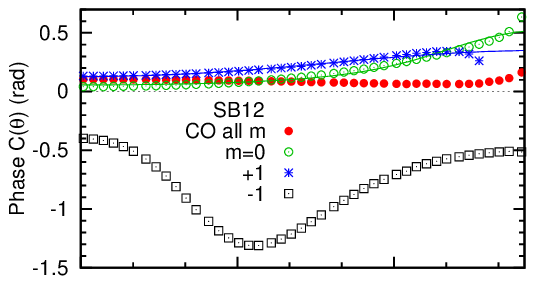}
\vs{-0.8cm}

\epsfxsize=6.5cm
\epsffile{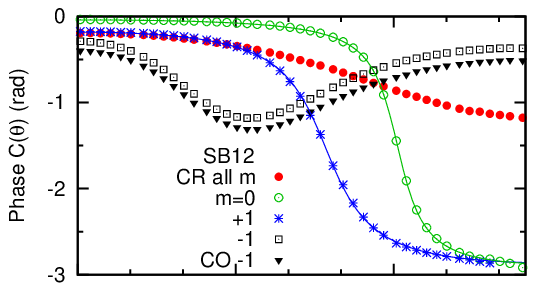}
\vs{-0.8cm}

\epsfxsize=6.5cm
\epsffile{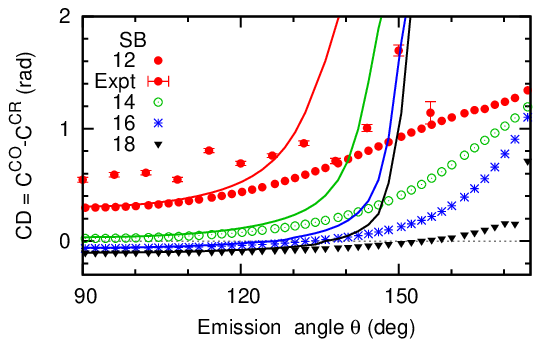}
\vs{-0.2cm}

\caption{ Angular dependent RABBITT phase ($C(\q)$ parameter) in SB12
  of argon obtained with CO (top panel) and CR (middle panel) circular
  polarization. The TDSE calculations with the sum over all $m$
  projections are shown with red dots whereas $m$ specific
  calculations are displayed with green circles ($m=0$), blue
  asterisks ($m=1$) and black squares ($m=-1$). The CO calculation
  with $m=-1$ is overplotted in the middle panel with balck triangles
  to demonstrate that $C_{m=-1}^{\rm CR} \simeq C_{m=-1}^{\rm CO}$. The
  parameterization with \Eref{CC} for $m=0$ and $m=+1$ is shown with
  the solid lines of the matching color.
Bottom: the CD$=C^{\rm CO} -C^{\rm CR}$ for
SB12-18 The dotted symbols and the solid lines of matching color
correspond to the all $m$ and $m=+1$ calculations, respectively.
The experimental CD values for SB12 are from \cite{HanMeng2023}.
\label{Fig3}}. 
\ms\ms
\end{figure}

\begin{figure}[h]
\epsfxsize=6.5cm
\epsffile{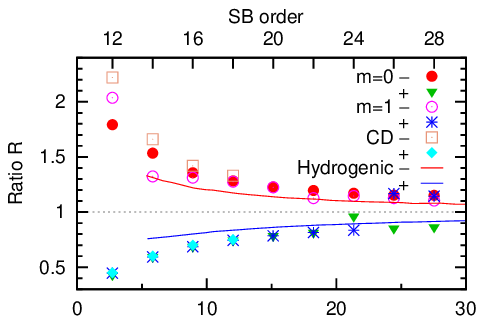}
\vs{-5mm}

\epsfxsize=6.5cm
\epsffile{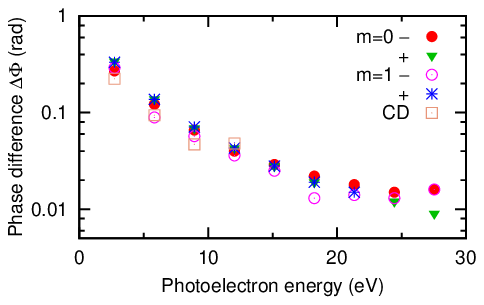}

\caption{Top: the ratios $R^\pm=|T_1^\pm/T_3^\pm|$ as extracted from
  the fit of \Eref{PP} to RABBITT   CO/CR phases with $m=0$
  and $m=+1$ as well the fit of the $m$-summed CD.  Comparison is made
  with a hydrogenic model employed in \cite{Boll2023}. Bottom: the
  analogous results for the phase differences $\Delta\Phi^\pm=\arg
  [T_1^\pm/T_3^\pm]$.
\label{Fig4}
\ms}
\end{figure}

\Eref{Mp}  defines the  RABBITT phases for $\phi=0$:
\ba
\label{PP}
C_{m=0}^{\rm CR/CO} &=& 
 \arg\Big[T_3^{-}T_3^{+*}\Big]
+ \arg\left[\bar P_3(\cos\q)-{T^{\pm}_1 \over T^{\pm}_3}\right]
\hs{5mm}
\\\nn
C_{m=+1}^{\rm CR/CO} &=& 
 \arg\Big[T_3^{-}T_3^{+*}\Big]
+ \arg\left[\frac16\bar P^1_3(\cos\q)-{T^{\pm}_1 \over T^{\pm}_3}\right]
\ea
We do not present here the explicit expressions for the RABBITT phases
in the $m=-1$ case for brevity. We  only note that it is entangled
with the $\pm$ amplitudes and that $C_{m=-1}^{\rm CR} =
C_{m=-1}^{\rm CO}$.
Eqs.~\eref{PP} offer a convenient parameterization of the RABBITT
phases for $m=0$ and $m=+1$ in terms of the ratios
$R^\pm=|T_1^\pm/T_3^\pm|$ and the phase differences
$\Delta\Phi^\pm=\arg [T_1^\pm/T_3^\pm]$. We demonstrate the utility of
this parameterization in \Fref{Fig3} where we plot the RABBITT phases
for the lowest SB12 in argon. The top and middle panels of this figure
display the CO and CR phases, respectively, for $m$-resolved and
$m$-summed cases. The ratios $R^\pm$ and phase differences
$\Delta\Phi^\pm$ are used as fitting parameters to fit the
corresponding CO/CR phases. These parameters are displayed in
\Fref{Fig4}. Two separate fits of $m=0$ and $m=+1$ produce the two
sets of parameters which should, in principle,  be identical. Their
actual difference serves as an accuracy indication of the fitting
procedure.  We note that, similarly to helium, $R^{+}<1<R^{-}$. The
only exception is $R^{+}_{m=0}$ which exceeds unity for highest
SB26-28. If this result is not accidental, it may signal a break up of
the Fano propensity rule due to the proximity to the Cooper minimum
where the discrete transition $3p\to Es$ gradually takes over the
$3p\to Ed$.

While the individual $m$ parameterization is very accurate and the
acquired sets of the ratios and phase differences are sufficiently
close between the $m=0$ and $m=+1$ projections, such a
parameterization is not practical experimentally. Indeed, a RABBITT
measurement on an unpolarized target atom corresponds to the
incoherent $m$ summation. To deduce the two-photon ionization
amplitude from such a measurement we analyze the CD exhibited in the
bottom panel of \Fref{Fig3}. Because of the CR/CO phase identity of
the $m=-1$ amplitude, its contribution vanishes from the CD. In
addition, the $m=0$ amplitude makes no contribution in the
polarization plane. Thus, close to $\q=90^\circ$, it is the $m=+1$
amplitude that brings the dominant contribution. This is illustrated
in the bottom panel of \Fref{Fig3} where the CD calculated from the
$m=+1$ amplitudes and the incoherent $m$ summation are exhibited. By
using the proximity of the two sets of results near $\q=90^\circ$, we
apply the $m=+1$ parameterization to the $m$-summed CD results over a
restricted angular range. Because the CD contains both the absorption
and emission $\pm$ amplitudes, the number of the fitting parameters
should be doubled in comparison with the separate CO and CR
fits. Such an extended fit becomes unstable and we need to impose
additional restrictions on the fitting parameters to improve its
accuracy. To do so, we require that $R^+=1/R^-$ and $\Delta\Phi^+ =
\Delta\Phi^-$. These restrictions are well justified as is seen from
the results of the separate CO/CR fits for $m=0$ and $m=+1$
projections. The results of the restricted $R$ and $\Delta\Phi$ fit of
the CD are overploted in \Fref{Fig4} for the lowest SB12-18 and are
found in fair agreement with the other four CR/CO and $m=0,+1$ sets of
parameters.

Similar to the helium results exhibited in \Fref{Fig2}, the hydrogenic
approximation to the amplitude ratios is more or less accurate for
sufficiently large photoelectron energies. However, close to the
threshold, the deviation between the absorption and emission $\pm$
ratios becomes significantly larger than predicted by the hydrogenic
model.

In conclusion, we devised a procedure to extract the complex
two-photon ionization amplitudes from the circular dichroic phase
acquired in a RABBITT measurement with circular polarized XUV and IR
pulses. Such measurements have been realized very recently by Han
{\em~et~al.} \cite{HanMeng2023,Han2023Separation} and demonstrated
distinct sets of RABBITT parameters with the co- and counter-rotating
XUV/IR radiations. In the case of helium and other $s$-electron
targets, the proposed method rests solely on the experimentally
accessible dichroic phase and does not require any further
approximations or simplifications. Moreover, as the amplitudes are
extracted from the angular dependent dichroic phase, the absolute
value of the latter is not needed. This is important experimentally as
this absolute value can be affected by the XUV harmonic group
delay. In the case of $p$-electron targets such as outer shells of
heavier noble gases, the amplitude extraction can be made fully
{\em~ab~initio} from the $m$-resolved dichroic phase. In this
procedure the proximity of the $m=0$ and $m=+1$ results serves as a
useful check of the accuracy of the method. In experimental
measurements on unpolarized targets, the two-photon amplitudes can be
extracted from the angular dependent CD  taken as the difference
between RABBITT phases acquired with the CO and CR circular
polarizations.  The CD determination of the two-photon ionization
amplitudes rests on the assumption that the absorption/emission ratio
and the phase difference are about the same in the CR and CO cases. This
assumption is well justified by the $m$-specific
tests of Ar. 

In a broader context, our results offer an opportunity to conduct a
complete experiment on the two-photon XUV+IR ionization whereupon the
moduli and phases of all the relevant ionization amplitudes are
determined experimentally. So far such experiments could only be
conducted in single-photon XUV ionization
\cite{Shigemasa1998,Wang2000}. An alternative method based on the
global fitting of the time- and angle-resolved RABBIT traces
\cite{Villeneuve2017} allows to extract the two-photon amplitudes in
various $m$-projected ionization channels \cite{HanMeng2023}. However,
these amplitudes are not independent and can be further reduced to the
most essential ``building blocks'' as demonstrated in the present
study. These blocks visualize very distinctly  the fundamental properties
of two-photon ionization such as the Fano propensity rule both for the
discrete \cite{PhysRevA.32.617} and the continuous transitions
\cite{Busto2019} as well as the emisson/absorption phase identity
\cite{Boll2023}. The proposed method also tests the validity of the
hydrogenic model \cite{Dahlstrom201353,Boll2023} which fails near the
threshold and in the vicinity of resonant excitations.

\paragraph*{Acknowledgment:}  
Resources of the National Computational Infrastructure facility have
been used in the present work.


\end{document}